\documentclass[preprint]{elsarticle}

\begin{document}

\title{ATP Hysteresis in Tripartite Synapses}

\author[rvt]{H. R. ~Noori\corref{cor1}}
\ead{hamid.reza.noori@iwr.uni-heidelberg.de, hnoori@princeton.edu}
\cortext[cor1]{Corresponding author}
\address[rvt]{Interdisciplinary Center for Scientific Computing, University of Heidelberg, \\ Im Neuenheimer Feld 294, 69120 Heidelberg, Germany}

\begin{abstract}
Recent experimental studies strongly suggest the influence of glial purinergic transmission in the modulation of synaptic dynamics. By releasing adenosine triphosphate (ATP), which accumulates as adenosine, astrocytes tonically suppressed synaptic transmission. The delayed multi-step feedback of the glial adenosine with the neuron suggest the existence of hysteresis phenomena, which are investigated in the present study from the theoretical point of view. The model suggests that a memory operator, tripartite synaptic plasticity, governs the mysterious delayed feedback inhibition caused by the action of adenosine on neuronal $A_1$ receptors and provides a powerful tool for further dynamical modeling tasks on tripartite synapses.     
\end{abstract}

\maketitle

\section{Introduction}

The nervous system consists of two classes of cell, the neuron and glia. According to the classical view of the nervous system, the numerically superior glial cells have inferior and passive roles in that they provide an ideal environment for neuronal-cell function. Recent evidence has demonstrated, however, that there is a dynamic reciprocal feedback communication between glia and neurons at the synapses (\cite{ara}, \cite{hay1}, \cite{new2}, \cite{vol}). Neurotransmitters released from presynaptic neurons evoke $Ca^{2+}$ concentration increases in perisynaptic glia. It is known that all types of glia (i.e. microglia, oligodendrocytes, astrocytes and Schwann cells) express a wide variety of neurotransmitter receptors (\cite{por}). Activated glia, then release neuroactive substances, called gliotransmitters, including glutamate and ATP. These gliotransmitters feed back on the one hand onto the presynaptic terminal either to facilitate or to suppress further release of neurotransmitter, and on the other hand stimulate directly postsynaptic neurons by producing either excitatory or inhibitory responses (\cite{fie}, \cite{hal}, \cite{ham}, \cite{hay2}, \cite{hay3}, \cite{pas}, \cite{zha}).  

Released ATP from glia could excite neurons directly through activation of P2X receptors. Alternatively, released ATP, once converted to adenosine, could inhibit neurons by activating $A_1$ receptors, as it does in the retina. Adenosine could also act presynaptically via $A_1$ and $A_2$ receptors to either depress or potentiate synaptic transmission (\cite{dun}, \cite{gor}, \cite{hal}, \cite{hay2}, \cite{kat}, \cite{mar}, \cite{mas}, \cite{new1}, \cite{new2}). 

It has been observed that during hypoxia, astrocytes contribute to regulate the excitatory synaptic transmission through the release of adenosine, which by acting on $A_1$ receptors reduces presynaptic neurotransmitter release (\cite{mar}).
Because adenosine is implicated in the control of wake-to-sleep transitions as well as responses to hypoxia, the identification of the role of the glia in regulating this transmitter offers mechanistic insights into these processes (\cite{bas}).

The kinetics of ATP hydrolysis and adenosine accumulation provides a framework with unique spatiotemporal conditions to control synaptic transmission. It takes $\approx 200 \ ms$ before adenosine begins to accumulate. The delays (pre-accumulation delay, accumulation delay) and different states of the action of glial ATP on neurons, suggest further investigations to reveal the mysterious nature of the delayed multi-state process (\cite{pas}). 

The aim of this article is the introduction of the concept of hysteresis in synaptic scale especially in the astrocytic ATP-related activities. This concept provide a simplified framework for the description of ATP kinetics that can be easily embedded into mathematical models dealing with neurochemistry of tripartite synapses. Thus, one can consider the present work as a tool for further investigations.

Based on Preisach model for hysteresis, nonideal relays $R^j=R_{a_j,b_j}$ are introduced. By $\mu_j$-weighted superposition of a finite bundle of relays the operator over the half plane $P$ of admissible thresholds, the Preisach operator $H_{\mu}$ is then constructed. Memory is represented by means of an antimonotone graph in $P$, which separates relays staying in different states. The reader is advised to refer to the appendix for more detailed mathematical description of hysteresis. 

It appears that this mathematical approach provides a realistic approximation for the kinetics of ATP and adenosine and their dynamical states, in terms of their action on presynaptic terminals. The memory, based on the delayed relay operator, characterizes faithfully this biological process.

\section{The Preisach Model}     

\bigskip\noindent
For the formulation of the ATP kinetics at tripartite synapses in the language of hysteresis, the relationship between ATP concentrations and activity behavior should be estabilished. ATP can be released in concentrations that might be subthreshold for the activation of $P2X$ receptors (micromolar) yet sufficiently high to allow an accumulation of adenosine that will activate $A1$ receptors (tens of nanomolar). 

\bigskip\noindent
This means, that the ATP kinetics can be decomposed into two phases: 

\begin{enumerate}
\item Hysteron $R^1=R_{a_1,b_1}$ of ATP concentrations acting on $P2X$ receptors. $a_1$ denotes the releasing point of ATP (concentration zero) and $b_1$ is representing the ATP concentration activating $P2$ purinergic receptors.
\item Hysteron $R^2=R_{a_2,b_2}$ of accumulated adenosine (hydrolized ATP) that acts inhibitor via $A1$ receptor activation. $a_2$ denotes the maximum concentration level of ATP  during pre-accumulation time of adenosine. $b_2$ is the accumulated adenosine concentration that is sufficient to activate $A1$ receptors. Here, the hydrolization of ATP to adenosine is seen as a change in the activity state of ATP;
\end{enumerate}

\bigskip\noindent
The parallel composition of these hysterons with equal weights $\mu_j$ leads to the finite Weiss-Preisach model of ATP hysteresis:

$$y(t) \ = \ \sum_{j=1}^2 \mu_j R^j[t_0,\eta_0] x(t)$$

where $x(t)$ is the concentration value of astrocytic ATP and $\eta_0$ the initial state of the system. This behavior is shown in figure 1.

\bigskip\noindent
Some important remarks should be mentioned at this point that are suggested by experimental data:

\begin{itemize}
\item $b_2 \approx 10^2 \cdot b_1$. This reflects the concentration thresholds for the activation of $A1$ (tens of nanomolar) and $P2$ receptors (micromolar).
\item $T_{b_2}-T_{b_1} \approx 200 ms$. $T_{b_j}$ denotes the instant correlating to a certain concentration. The difference is in accordance with the pre-accumulation time of adenosine. 
\end{itemize}

\begin{figure}[h]
\begin{center}
\includegraphics[width=\textwidth]{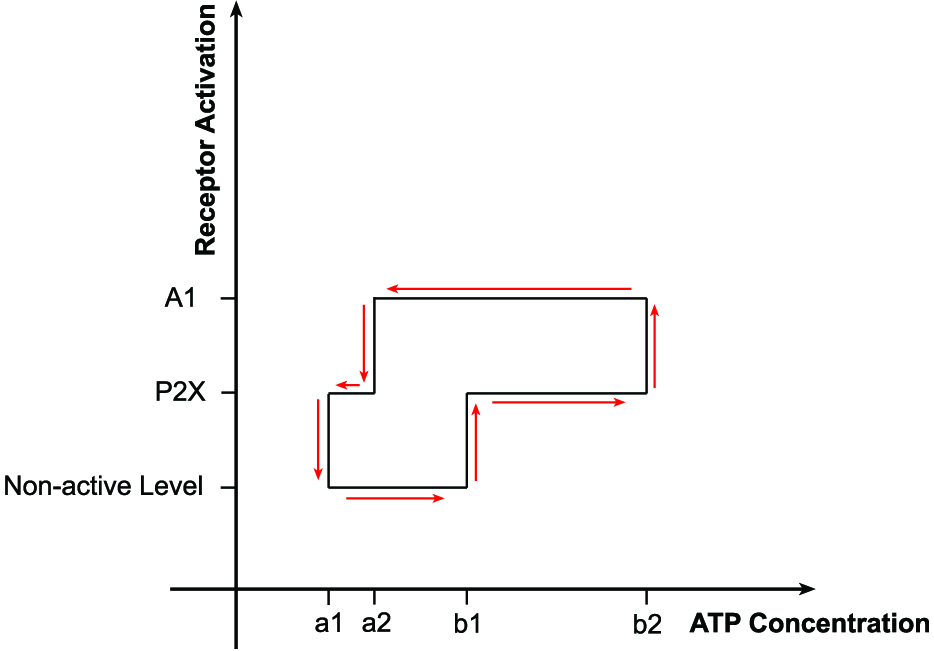}
\vspace{2.5cm}
\caption{The hysteresis diagramm of the ATP concentration and purinergic receptor activation. Two hysterons are parallel connected: First hysteron $R_{a_1,b_1}$ is the delay relay operator of the accumulated to adenosine hydrolized ATP concentration; The second hysteron $R_{a_2,b_2}$, is the delay relay operator of the $P2$ activating ATP concentrations. The extremal values are denoting the concentration threshold of the release point of ATP and the maximum concentration of tripartite synaptic ATP.}
\label{single}
\end{center}
\end{figure}  

\bigskip\noindent
This simple formulation output variable $y(t)$ of the two-phaseal ATP-induced activity provide an useful framework for investigating the purinergic role in neuron-glia interactions at tripartite synapses. It also implies that $y(t)$ as a hysteresis process, is governed by rate independent memories of ATP concentrations at different instants.

\section{Discussion}

\bigskip\noindent
Although this concept is a rough simplification of the ATP-related processes and does not provide new insights in the physiology of the tripartite synapses at this level, it supplies a powerful tool for modeling of synaptic behavior. It should be considered as a pre-step of the modeling the neurochemical interaction in tripartite synapses, roughly spoken as a physical characterization that provides a useful mathematical relation for further dynamical modeling tasks. The Memory of ATP concentrations determines the kinetics of purinergic activity at tripartite synapses. The knowledge about this biological memory could be of use for
physiological investigations. By enlarging the number of hysterons e.g. within the knowledge of the release procedure of astrocytic ATP, in terms of vesicular or non-vesicular release, the models refinement and accuracy increases that can lead to specific applications of  its results other than its usage as a completion tool for other models. 

\bigskip\noindent

\paragraph{Acknowledgment.}
The financial support by the Interdisciplinary Center for Scientific Computing and the International Graduiertenkolleg 710 (DFG) and the funding by National Genome Research Network are acknowledged.  

\bigskip\noindent

\paragraph{Appendix}

The aim of this appendix, which is merely based on the work of Krasnosel'skii and Pokrovskii (\cite{kra}), and Visintin (\cite{vis}), is to introduce the general idea of the Preisach model of hysteresis to unexperienced readers.
 
The hysteresis nonlinearities are defined as deterministic, rate independent operators. The standard approach to hysteresis, is the introduction of elementary hysteresis nonlinearities, called hysterons, and treatment of complex hysteresis processes as superposition (parallel connection) of these hysterons. Among hysterons, the most important are probably the nonideal relays, or thermostat nonlinearities. 

The nonideal relay (with the threshold values $a < b$) is the simplest hysteretic transducer. This transducer is denoted by $R_{a,b}$. Its output $y(t)$ can take one of the two values $0$ or $1$: at any moment the relay is either 'switched off' or 'switched on'.
The output variable $y(t)$
$$y(t) \ = \ R_{a,b}[t_0,\eta_0]x(t) \ , \ t \geq t_0,$$
depends on the input variable $x(t) \in C^0([t_0,T])$ and on the initial state $\eta_0 \in \{0 , 1 \}$. The values of the output function $y(t)$ at an instant $t$ are defined by the following explicit formula:
\begin{itemize}
\item $\eta_0$, if $a < x(\tau) < b$ $\forall \ \tau \in [t_0,t]$; 
\item $1$, if $\exists \ t_1 \in [t_0,t]$: $x(t_1) \geq b$, $x(\tau) > a$ $\forall \ \tau \in [t_1,t]$; 
\item $0$, if $\exists \ t_1 \in [t_0,t]$: $x(t_1) \leq a$, $x(\tau) < b$ $\forall \ \tau \in [t_1,t]$;
\end{itemize}

Consider a family $R$ of relays $R^w=R_{a_w,b_w}$ with threshold values $a_w,b_w$ (a bundle of relays), with $w \in \Omega$. The index set $\Omega$ may be finite or infinite. Suppose that the set $\Omega$ is endowed with a Borel measure $\mu$. Further suppose that the functions $a_w,b_w$ are measurable with respect to $\mu$. We call any measurable function $\eta(w): \Omega \rightarrow \{0,1\}$ the initial state of the bundle $R$.

For any initial state $\eta_0(w)$ and any continuous input $x(t)$, $t \geq t_0$, we define the Preisach hysteresis operator as the integral operator:
$$y(t) \ = \ H_{\mu}[t_0,\eta_0] x(t) \ = \ \int_{\Omega} R^w[t_0, \eta_0(w)] x(t) d \mu \ , \ t \geq t_0 \ .$$

\end{document}